\documentclass[twocolumn,tighten]{aastex62}

\usepackage{graphics}
\usepackage{amsmath}
	
\newcommand{\lea}{{\>\rlap{\raise2pt\hbox{$<$}}\lower3pt\hbox{$\sim$} \>}}
\newcommand{\gea}{{\>\rlap{\raise2pt\hbox{$>$}}\lower3pt\hbox{$\sim$} \>}}

\def\hi{\mbox{H\,{\sc i}}}

\shorttitle{Angular momentum and galaxy formation}
\shortauthors{Fall \& Romanowsky}

\begin{document}

\title{ ANGULAR MOMENTUM AND GALAXY FORMATION REVISITED:  \\
          SCALING RELATIONS FOR DISKS AND BULGES }

\author{S. Michael Fall}
\affiliation{Space Telescope Science Institute,
         3700 San Martin Drive, Baltimore, MD 21218, USA}
\author{Aaron J. Romanowsky}
\affiliation{Department of Physics \& Astronomy, 
San Jos{\'e} State University,  One Washington 
Square, San Jose, CA 95192, USA}
\affiliation{University of California Observatories,
1156 High Street, Santa Cruz, CA 95064, USA}

\begin{abstract}

We show that the stellar specific angular momentum $j_{\star}$, mass $M_{\star}$, and bulge fraction $\beta_{\star}$ of normal galaxies of all morphological types are consistent 
with a simple model based on a linear superposition of independent disks and bulges.
In this model, disks and bulges follow scaling relations of the form $j_{\star{\rm d}} \propto M_{\star{\rm d}}^{\alpha}$ and $j_{\star{\rm b}} \propto M_{\star{\rm b}}^{\alpha}$ with $\alpha = 0.67 \pm 0.07$ but offset from each other by a factor of $8 \pm 2$ over the mass range $8.9 \leq \log (M_{\star}/M_{\odot}) \leq 11.8$.
Separate fits for disks and bulges alone give $\alpha = 0.58 \pm 0.10$ and  $\alpha = 0.83 \pm 0.16$, respectively.
This model correctly predicts that galaxies follow a curved 2D surface in the 3D space of $\log j_{\star}$, $\log M_{\star}$, and $\beta_\star$.
We find no statistically significant indication that galaxies with classical and pseudo bulges follow different relations in this space, although some differences are permitted within the observed scatter and the inherent uncertainties in decomposing galaxies into disks and bulges.
As a byproduct of this analysis, we show that the $j_{\star}$--$M_{\star}$ scaling relations for disk-dominated galaxies from several previous studies are in excellent agreement with each other.
In addition, we resolve some conflicting claims about the $\beta_\star$-dependence of the $j_{\star}$--$M_{\star}$ scaling relations.
The results presented here reinforce and extend our earlier suggestion that the distribution of galaxies with different $\beta_{\star}$ in the $j_{\star}$--$M_{\star}$ diagram constitutes an
objective, physically motivated alternative to subjective classification schemes such as the Hubble sequence.

\end{abstract}


\section{Introduction}

Specific angular momentum ($j = J/M$) and mass ($M$) are two of the most basic properties of galaxies.
Together, they largely determine another basic property---characteristic size (such as half-mass radius $R_{\rm h}$)---especially for disk-dominated galaxies. 
Thus, the correlation between $j$ and $M$ constitutes one of the most fundamental scaling relations for galaxies, as important as those between rotation velocity, velocity dispersion, characteristic size, and mass. 
We have studied the galactic $j$--$M$ relation from both observational and theoretical perspectives (Fall 1983; Romanowsky \& Fall 2012; Fall \& Romanowsky 2013, hereafter Papers 0, 1, and 2).
The present paper is a continuation of this series.
In the following, when relevant, we distinguish between the stellar, baryonic, and halo parts of galaxies with the subscripts $\star$, bary, and halo, and between their disk and bulge components with the subscripts d and b. 

We have found that both disk-dominated galaxies and bulge-dominated galaxies (mainly ellipticals) obey power-law scaling relations of the form $j_{\star} \propto M_{\star}^{\alpha}$ with essentially the same exponent, $\alpha = 0.6 \pm 0.1$, and normalizations that differ by a factor of $\sim 5$. 
Our results are based on a sample in which most galaxies have classical bulges (genuine spheroids) rather than pseudo bulges (disk-like structures).
In a plot of $\log j_{\star}$ against $\log M_{\star}$, galaxies of different morphological type and bulge fraction $\beta_{\star} \equiv M_{{\star}{\rm b}} / ( M_{{\star}{\rm d}} +  M_{{\star}{\rm b}} )$ follow nearly parallel relations, filling the region between the sequences of disk-dominated and bulge-dominated galaxies.   
Based on this finding, we have proposed that the distribution of galaxies with different $\beta_\star$ in the $j_{\star}$--$M_{\star}$ diagram constitutes an objective, physically motivated alternative to subjective classification schemes such as the Hubble sequence. 

The parallel sequences of galaxies of different bulge fraction in the $j_{\star}$--$M_{\star}$ diagram suggest a picture in which galactic disks and spheroids are essentially independent objects, formed by distinct physical processes. 
Disks likely formed relatively quiescently by diffuse gas settling within dark-matter halos, 
while spheroids likely formed more violently by colliding streams and clumps of cold gas and by merging of smaller galaxies.
Disk-dominated galaxies are those in which major mergers played little or no role, while spheroid-dominated galaxies either never acquired a substantial disk or else acquired one and later lost it by stripping or merging.
In this picture, most normal galaxies may be regarded, in a first approximation, as a linear superposition of a flat disk and a round spheroid, each of which lies along the corresponding $j_{\star}$--$M_{\star}$ sequence.  
The primary purpose of this paper is to make a quantitative test of this picture.  

The observed $j_{\star}$--$M_{\star}$ scaling relations also link well with galaxy-formation theory.  
The galactic halos that form by hierarchical clustering in a dark-matter dominated universe (such as $\Lambda$CDM) obey the scaling relation $j_{\rm halo} \propto M_{\rm halo}^{\alpha}$ with $\alpha = 2/3$, an exponent remarkably similar to that for the stellar parts of galaxies. 
The halo scaling relation follows directly from the fact that the spin parameter $\lambda_{\rm halo}$ and mean internal density $\bar\rho_{\rm halo}$ are independent of $M_{\rm halo}$.
By comparing the $j_{\star}$--$M_{\star}$ and $j_{\rm halo}$--$M_{\rm halo}$ relations, mediated by an $M_{\star}$--$M_{\rm halo}$ relation, we found that galactic disks have a fraction of specific angular momentum relative to their surrounding halos of $f_j \equiv j / j_{\rm halo} \approx 0.8$, while galactic spheroids have a fraction $f_j \approx 0.15$. 
The first of these agrees well with the postulated value $f_j \approx 1$ in simple analytical models of galactic disk formation (Fall \& Efstathiou 1980; Dalcanton et al.\ 1997; Mo et al.\ 1998). 

The $j$--$M$ scaling relations have been the focus of further observational study, both for low-redshift galaxies (Obreschkow \& Glazebrook 2014; Cortese et al.\ 2016; Butler et al.\ 2017; Chowdhury \& Chengalur 2017; Elson 2017; Kurapati et al.\ 2018; Lapi et al.\ 2018b; Posti et al.\ 2018a; Rizzo et al.\ 2018; Sweet et al.\ 2018) and for high-redshift galaxies (Burkert et al.\ 2016; Contini et al.\ 2016; Harrison et al.\ 2017; Shi et al.\ 2017; Swinbank et al.\ 2017; Tadaki et al.\ 2017; Alcorn et al.\ 2018).  
Several recent studies have examined the relation between galaxy sizes and halo sizes, a corollary of the $j$--$M$ relation (Kravtsov 2013; Kawamata et al.\ 2015; Shibuya et al.\ 2015; Huang et al.\ 2017; Kawamata et al.\ 2018; Okamura et al.\ 2018).  
The $j$--$M$ scaling relations have been a benchmark for some recent analytical and semi-analytical models (Stevens et al.\ 2016; Shi et al.\ 2017; Lapi et al.\ 2018a; Posti et al.\ 2018b; Zoldan et al.\ 2018).  
They have also been the targets of many recent hydrodynamical simulations, some with large volume but relatively low resolution (Genel et al. 2015; Pedrosa \& Tissera 2015; Teklu et al.\ 2015; Zavala et al.\ 2016; DeFelippis et al.\ 2017; Lagos et al.\ 2017; Stevens et al.\ 2017; Lagos et al.\ 2018; Schulze et al.\ 2018) and others with small volume (zoom-in) but higher resolution (Agertz \& Kravtsov 2016; Grand et al.\ 2017; Soko{\l}owska et al.\ 2017; El-Badry et al.\ 2018; Obreja et al.\ 2018). 

The studies cited above generally confirm our $j_{\star}$--$M_{\star}$ scaling relations, particularly the exponent $\alpha \approx 0.6$ for disk-dominated galaxies. 
The exceptions to this near-consensus are the works by Obreschkow \& Glazebrook (2014) and Sweet et al.\ (2018), which found $\alpha \approx 1.0$ for galaxies of the same bulge fraction, including $\beta_{\star} = 0$.
Obreschkow \& Glazebrook (2014) interpreted this to mean that the angular momenta of galactic disks are influenced in some way by the prominence of galactic bulges, i.e., that these two components are not independent, in contradiction to the picture discussed above.
Complicating this comparison, however, is the fact that most of the galaxies in the Obreschkow \& Glazebrook (2014) and Sweet et al.\ (2018) samples have pseudo bulges rather than classical bulges. 
Thus, a secondary purpose of this paper is to resolve the apparent discrepancy between their work and ours.   

The remainder of this paper is organized as follows.
In Section~2, we compare and contrast four determinations of the $j_\star$--$M_\star$ relation for galaxies of different bulge fraction, revealing some important similarities and differences.
In Section~3, we present the corresponding two-dimensional (2D) surfaces defined by these relations in the three-dimensional (3D) space of $j_\star$, $M_\star$, and $\beta_\star$, and show that they are consistent with our picture of independent disks and spheroids.
We summarize our results and discuss their implications in Section~4. 
We make some detailed comparisons between our estimates of $j_\star$, $M_\star$, and $\beta_\star$ and those of others in the Appendix.

\section{2D Relations Between  \MakeLowercase{\it J}$_{\star}$ {\rm and} $M_{\star}$}

Before we consider the distribution of galaxies in the 3D space of specific angular momentum $j_\star$, mass $M_\star$, and bulge fraction $\beta_\star$, it is helpful to review several determinations of the 2D scaling relations between $j_{\star}$ and $M_{\star}$ for galaxies in different ranges of $\beta_\star$.
In particular, we focus on the results from our work (Paper~2), Obreschkow \& Glazebrook (2014), Posti et al.\ (2018a), and Sweet et al.\ (2018).
These four $j_{\star}$--$M_{\star}$ scaling relations are plotted here in Figure~1, with galaxies of different $\beta_\star$ represented by symbols of different colors and shapes. 
Evidently, there are some important similarities and differences between them (Section~2.2). 
In order to understand why they agree in some respects and disagree in others, we briefly review some of the key assumptions and procedures involved in their derivations  (Section~2.1). 
For a more complete description of how these $j_{\star}$--$M_{\star}$ scaling relations were derived, we refer interested readers to the original papers. 

\begin{figure*}
\includegraphics[width=\textwidth]{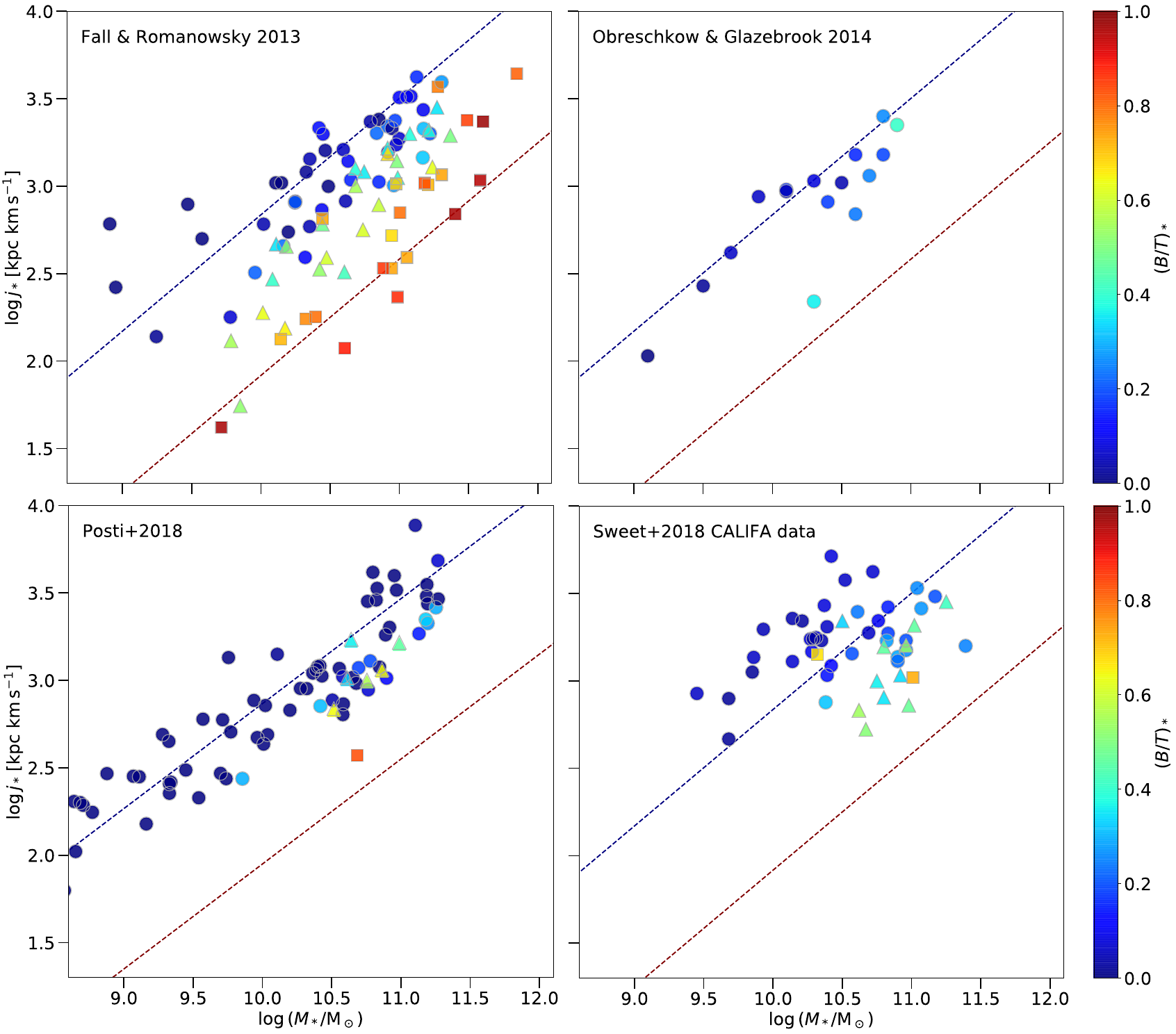}
\caption{
Stellar specific angular momentum $j_{\star}$ plotted against stellar mass $M_{\star}$ for galaxies of different stellar bulge fraction, $\beta_\star \equiv (B/T)_{\star}$, from the references indicated in the upper-left corners of the panels.
The colors and shapes of the plotted symbols indicate the bulge fractions, with circles for $0 \leq \beta_\star < 1/3$, triangles for $1/3 \leq \beta_\star < 2/3$, and squares for $2/3 \leq \beta_\star \leq 1$.  
In all four panels, the parallel dashed lines represent the $j_{\star}$--$M_{\star}$ scaling relations for disks and bulges derived from the 3D fits of Equations (2) and (3) to our dataset as described in Section~3.   
Note that the scaling relations for disk-dominated galaxies from our work, Obreschkow \& Glazebrook (2014), and Posti et al.\ (2018a) are in excellent agreement with each other, while that from Sweet et al.\ (2018) is offset from the others.  
Note also that these samples, with the exception of ours, are largely devoid of bulge-dominated galaxies.
}
\end{figure*}

\subsection{Samples and Methods}

Our $j_{\star}$--$M_{\star}$ relation is based on a sample of 94 galaxies:  57 spirals, 14 lenticulars, and 23 ellipticals.\footnote
{This sample is the same as the one in Paper~2 (despite a misprint there) but differs from the one in Tables 3--5 of Paper~1 because 11 galaxies lack color information and two are peculiar.}
This sample spans a wide range of mass, $8.9 \leq \log ( M_{\star} /M_{\odot} ) \leq 11.8$, and the full range of bulge fraction, $0 \leq \beta_\star \leq 1.0$, with median $\beta_\star = 0.35$.
The main selection criterion for this sample was the availability of photometric and kinematic data extending to large radii:  $\sim 2 R_{\rm e}$ in all cases and up to $\sim 10 R_{\rm e}$ in some cases (where $R_{\rm e}$ is the effective or projected half-light radius). 
The large radial extent of the data has allowed us to obtain convergent estimates of $j_{\star}$ and $M_{\star}$ with relatively little extrapolation beyond the outermost measurements (see Section~3 of Paper~1).
Most of the surface brightness profiles of the disks and bulges of spiral galaxies come from Kent (1986, 1987, 1988).
His method of decomposition matches the 2D images of galaxies with combinations of disks and bulges with pre-specified 3D shapes (flat and round) rather than pre-specified surface brightness profiles (exponential and S{\'e}rsic).
We estimated the bulge fractions of lenticular and elliptical galaxies from the observed ratios of stellar rotation velocities and velocity dispersions ($v_\star/\sigma_\star$), as calibrated by photometric decompositions (see Appendix~D of Paper~1).

For spiral galaxies, we estimated the specific angular momentum and mass separately for the disk and bulge components, namely ($j_{{\star}{\rm d}}$, $j_{{\star}{\rm b}}$) and ($M_{{\star}{\rm d}}$, $M_{{\star}{\rm b}}$).
This was necessary because these disks and bulges have distinct mass-to-light ratios, indicated by their colors, and distinct kinematics.
For each disk, we derived $j_{{\star}{\rm d}}$ directly from the surface brightness profile and the H$\alpha$ or \hi\ rotation curve (or approximations to them), while for each bulge we derived $j_{{\star}{\rm b}}$ indirectly from the surface brightness profile and the rotation velocity $v_{{\star}{\rm b}}$ estimated from the velocity dispersion $\sigma_{{\star}{\rm b}}$ and ellipticity $\epsilon_{\rm b}$ and the mean relation between $v_{{\star}{\rm b}} /\sigma_{{\star}{\rm b}}$ and $\epsilon_{\rm b}$ for the bulges of similar galaxies.
We then combined our estimates of specific angular momentum and mass for each component into the corresponding totals, $j_{\star} = ( j_{{\star}{\rm d}} M_{{\star}{\rm d}} + j_{{\star}{\rm b}} M_{{\star}{\rm b}} )/ M_{\star}$ and $M_{\star} = M_{{\star}{\rm d}} + M_{{\star}{\rm b}}$, for each spiral galaxy.
For lenticular and elliptical galaxies, which have disks and bulges of similar mass-to-light ratio, we estimated the total values $j_{\star}$ and $M_{\star}$ directly from the overall surface brightness profiles and stellar rotation profiles derived from optical absorption-line spectra and globular cluster and planetary nebula velocities. 

In Paper~1, we estimated the stellar masses of both disks and bulges from their $K$-band (2.2 ${\mu}$m) luminosities and an assumed universal mass-to-light ratio, $M_{\star}/L_K = 1.0$.
Subsequently, in Paper~2, we revised our mass estimates for disks and bulges based on their observed $B-V$ colors and the predicted relation between $M_{\star}/L_K$ and $B-V$ from stellar population models with different star formation histories.
These revisions are fairly modest, with typical values $(M_{\star}/L_K)_{\rm d} \approx 0.5$ for the disks of late-type spirals, and $(M_{\star}/L_K)_{\rm b} \approx 0.8$ for all bulges.
The revised mass-to-light ratios do not affect the specific angular momenta of the disk and bulge components derived in Paper~1, but they do affect the total values $j_{\star}$, as these are mass-weighted sums of $j_{{\star}{\rm d}}$ and $j_{{\star}{\rm b}}$. 
We used, but did not publish, the resulting estimates of $j_{\star}$, $M_{\star}$, and $\beta_\star$ in Paper~2; we list them here in Table~1 and plot them in Figure~1.

The Obreschkow \& Glazebrook (2014) $j_{\star}$--$M_{\star}$ relation is based on 16 late-type spiral galaxies in the THINGS\footnote{THINGS is an acronym for The HI Nearby Galaxy Survey.} survey with {\it Spitzer} 3.6 ${\mu}$m surface photometry and \hi\ rotation curves from Leroy et al.\ (2008).
This sample spans a moderate range of mass, $9.1 \leq \log ( M_{\star} /M_{\odot} ) \leq 10.9$, but only a narrow range of bulge fraction, $0 \leq \beta_\star \leq 0.3$, with median $\beta_\star = 0.10$.
Obreschkow \& Glazebrook (2014) decomposed these galaxies by fitting a parametric model with an exponential disk and a S{\'e}rsic  bulge to the observed surface brightness profiles.
The masses of the galaxies were estimated using an empirical conversion between 3.6 ${\mu}$m and $K$-band luminosities and an assumed universal mass-to-light ratio, $M_{\star}/L_K = 0.5$, for both disks and bulges.
The specific angular momenta were estimated from the \hi\ rotation curves, assuming that the stellar disks and bulges corotate with each other and with the \hi.
Obreschkow \& Glazebrook (2014) derived both the stellar and baryonic $j$--$M$ scaling relations for this sample (including both stars and cold gas in the latter) and found that they were remarkably similar.
For direct comparison with the other $j_{\star}$--$M_{\star}$ relations, we plot only their stellar relation in Figure~1. 

The Posti et al.\ (2018a) $j_{\star}$--$M_{\star}$ relation is based on 92 spiral galaxies, mostly of late type, in the SPARC\footnote{SPARC is an acronym for Spitzer Photometry and Accurate Rotation Curves.} survey with {\it Spitzer} 3.6 ${\mu}$m surface photometry and \hi\ rotation curves from Lelli et al.\ (2016).
This sample spans an exceptionally large range of mass, $7.0 \leq \log ( M_{\star} /M_{\odot} ) \leq 11.3$, and a narrow range of bulge fraction, with $0 \leq \beta_\star \leq 0.3$ for 90\% of the galaxies.
Posti et al.\ (2018a) adopted the decompositions of these galaxies from Lelli et al.\ (2016), who fitted a non-parametric model consisting of an unspecified bulge and inner disk and an exponential outer disk to the observed surface brightness profiles.
The masses of the disks and bulges were estimated from their 3.6 ${\mu}$m luminosities and the adopted mass-to-light ratios $(M_{\star}/L_{[3.6]})_{\rm d} = 0.5$ and $(M_{\star}/L_{[3.6]})_{\rm b} = 0.7$, corresponding to $(M_{\star}/L_K)_{\rm d} \approx 0.4$ and $(M_{\star}/L_K)_{\rm b} \approx 0.6$.
Posti et al.\ (2018a) estimated the specific angular momenta from the \hi\ rotation curves, assuming that the stellar disks and bulges corotate with each other but with a small lag (asymmetric drift) relative to the \hi. 

The $j_{\star}$--$M_{\star}$ relation published by Sweet et al.\ (2018) is based on 50 galaxies in the CALIFA\footnote{CALIFA is an acronym for Calar Altar Legacy Integral Field Area.} survey, 16 galaxies in the THINGS survey, and 25 galaxies in our sample, with $j_{\star}$ and $M_{\star}$ estimated by different methods for each of these subsamples.
For galaxies in the THINGS survey and in our sample, the values of $j_{\star}$ and $M_{\star}$ were taken directly or adapted from Obreschkow \& Glazebrook (2014) and from our Paper~1, respectively.
Here, we consider only the CALIFA part of the Sweet et al.\ (2018) $j_{\star}$--$M_{\star}$ relation, in order to make meaningful comparisons with the other $j_{\star}$--$M_{\star}$ relations derived by different authors from independent samples of galaxies.

The CALIFA part of the Sweet et al.\ (2018) $j_{\star}$--$M_{\star}$ relation is based on galaxies of all morphological types except diskless ellipticals, with optical surface photometry from M\'endez-Abreu et al.\ (2017) and integral-field spectroscopy from Falc\'on-Barroso et al.\ (2017).   
This sample spans a moderate range of mass, $9.5 \leq \log ( M_{\star} /M_{\odot} ) \leq 11.4$, and a moderate range of bulge fraction, $0 \leq \beta_\star \leq 0.7$, with median $\beta_\star = 0.2$.
Sweet et al.\ (2018) adopted the decompositions of these galaxies from M\'endez-Abreu et al.\ (2017), who fitted a parametric model with an exponential disk (with a possible upward or downward outer bend) and a S{\'e}rsic  bulge to the observed 2D isophotes. 
The total masses $M_{\star}$, taken from Falc\'on-Barroso et al.\ (2017), were estimated from total luminosities in several bands and mass-to-light ratios predicted by stellar population models that matched the observed colors (as described by Walcher et al.\ 2014).
Sweet et al.\ (2018) estimated the total specific angular momenta $j_{\star}$ from the projected density and velocity maps derived from the surface photometry and integral-field spectroscopy of stellar absorption lines, assuming that all stars move on coplanar circular orbits (with no velocity dispersion).  

In summary, for spiral galaxies, our study and those of Obreschkow \& Glazebrook (2014) and Posti et al.\ (2018a) adopted similar methods for estimating the disk contributions to $j_\star$ (from H$\alpha$ and \hi\ rotation curves) and $M_\star$ (from near-IR luminosities and similar mass-to-light ratios).
However, these studies differed substantially in their treatment of bulges: first, in the methods of disk--bulge decomposition, and second in the assumptions about whether or not disks and bulges have the same rotation velocities.
The simplifying assumption that disks and bulges corotate, made by Obreschkow \& Glazebrook (2014) and Posti et al.\ (2018a), leads to acceptably small errors in $j_\star$ for disk-dominated galaxies but not for bulge-dominated galaxies.
Since our study aimed to derive the $j_{\star}$--$M_{\star}$ relation over the full range of bulge fractions, we estimated the bulge contributions to $j_\star$ for spiral galaxies indirectly from the velocity dispersion and ellipticity of their bulges, independently of the H$\alpha$ and \hi\ rotation curves of their disks, and the total $j_\star$ for lenticular and elliptical galaxies directly from the stellar rotation profiles. 

Based on the tests described in Paper 1, we estimate the following typical errors: $\varepsilon(\log j_\star) \approx 0.15$, $\varepsilon(\log M_\star) \approx 0.10$, and $\varepsilon(\beta_\star) \approx 0.10$ for spiral galaxies and $\varepsilon(\log j_\star) \approx 0.20$, $\varepsilon(\log M_\star) \approx 0.10$, and $\varepsilon(\beta_\star) \approx 0.20$ for lenticular and elliptical galaxies.
These are meant to include all sources of uncertainty and thus to represent {\it total} errors.
In particular, they include uncertainties in radial extrapolations of photometric and kinematic data, inclination angles, mass-to-light ratios, and distances.
They also include the inevitable deviations of real galaxies from the idealizations required to decompose them into disks and bulges (either pre-specified 3D shapes or surface-brightness profiles) and from the assumptions about bulge rotation.
In comparison with these uncertainties, measurement errors are usually negligible.
The errors in $\log j_\star$, $\log M_\star$, and $\beta_\star$ quoted by Obreschkow \& Glazebrook (2014), Posti et al.\ (2018a), and Sweet et al. (2018) are smaller than our estimates because they exclude one or more sources of uncertainty mentioned above and therefore represent {\it partial} errors. (See Section~2.2 and the Appendix for further discussion of the errors.)

\subsection{Comparison of Results}

Figure~1 reveals some interesting similarities and differences between the results from these four studies.  
The first conclusion apparent from Figure~1 is that the $j_{\star}$--$M_{\star}$ relations for disk-dominated galaxies from our Paper~2, Obreschkow \& Glazebrook (2014), and Posti et al.\ (2018a) agree remarkably well with each other.
We quantify this impression by fitting a power-law model in the form
\begin{equation}
\log (j_\star/j_0) = \alpha \log (M_\star/M_0) ,
\end{equation} 
with $\log (M_0/M_\odot) = 10.5$,
by least-squares minimization in the $j_\star$ direction and bootstrap uncertainty analysis, over the mass range $9.5 \leq \log (M_\star/M_\odot) \leq 11.5$.
Restricting the fits to galaxies with $\beta_\star \leq 0.1$, i.e., essentially pure disks (within the uncertainties in $\beta_\star$), we find $\alpha = 0.58\pm0.10$, $\log j_0 = 3.07\pm0.03$, and $\sigma (\log j_\star) = 0.16$ for our dataset, $\alpha = 0.63 \pm 0.08$, $\log j_0 = 3.16 \pm 0.04$, and $\sigma (\log j_\star) = 0.09$ for the Obreschkow \& Glazebrook (2014) dataset, and $\alpha = 0.61 \pm 0.06$, $\log j_0 = 3.10 \pm 0.03$, and $\sigma (\log j_\star) = 0.19$ for the Posti et al.\ (2018a) dataset (with $j_0$ expressed in units of kpc km s$^{-1}$).

These $j_{\star}$--$M_{\star}$ relations are virtually identical within the statistical errors, both in exponent ($\alpha \approx 0.6$) and normalization ($\log j_0 \approx 3.1$).
The dispersions of individual points about the mean relations in the vertical direction are also similar ($\sigma (\log j_\star) \approx$~0.1--0.2) and roughly consistent with the corresponding typical error $\varepsilon(\log j_\star)$.
We make some further comparisons between our dataset and those of Obreschkow \& Glazebrook (2014) and Posti et al.\ (2018a) in the Appendix.
In particular, we compare the independent estimates of $j_{\star}$, $M_{\star}$, and $\beta_{\star}$ for the 6--10 galaxies in common between these samples. 
The mean offsets are small and the dispersions are $\sigma(\log j_\star) = 0.11$, $\sigma(\log M_\star) = 0.10$, and $\sigma(\beta_\star) = 0.09$, again roughly consistent with the corresponding typical errors $\varepsilon(\log j_\star)$, $\varepsilon(\log M_\star)$, and $\varepsilon(\beta_\star)$.
These comparisons indicate that the estimated total errors quoted at the end of Section 2.1 are approximately correct for all three datasets.

There is a simple reason for the excellent agreement between different $j_{\star}$--$M_{\star}$ relations for disk-dominated galaxies.
Most galactic disks are similar to each other over a wide radial range, with exponential surface density profiles and flat rotation curves, characterized by the radial scale $R_{\rm d}$ and the rotation velocity $V_{\rm f}$, respectively.
In the inner regions, the surface density profiles and rotation curves vary much more among galaxies, but these variations have little influence on the values of $j_{{\star}{\rm d}}$ and $M_{{\star}{\rm d}}$.  
Thus, the relation, $j_{{\star}{\rm d}} = 2 R_{\rm d} V_{\rm f}$, for an ideal disk with an exponential surface density profile and a flat rotation curve, is a good approximation for most real disks of giant spiral galaxies.
This simplification makes the $j_{\star}$--$M_{\star}$ relation for disk-dominated galaxies relatively easy to determine.
Indeed, it has not changed much since the original derivation 35 years ago (Paper~0).
The more difficult task is to determine the $j_{\star}$--$M_{\star}$ relation for bulge-dominated galaxies.

The second conclusion apparent from Figure~1 is that there is a systematic offset between the Sweet et al.\ (2018) $j_{\star}$--$M_{\star}$ relation for disk-dominated CALIFA galaxies and the other three $j_{\star}$--$M_{\star}$ relations for disk-dominated galaxies.
When we fit Equation (1) to the galaxies with $\beta_\star \leq 0.1$ in the Sweet et al.\ (2018) CALIFA sample, we obtain $\alpha = 0.56 \pm 0.14$ and $\log j_0 = 3.41 \pm 0.05$, essentially the same exponent as the other $j_{\star}$--$M_{\star}$ relations, but a higher normalization by a factor of 2.0. 
Sweet et al.\ (2018) do not mention this offset in their paper.
We suspect it arises from errors in their calculations of specific angular momentum, as discussed in the Appendix.
In any case, the large, unexplained offset introduces a serious bias in the combined $j_{\star}$--$M_{\star}$ relation Sweet et al.\ (2018) derived from the CALIFA, THINGS, and our Paper~1 datasets.  
This offset is one of the reasons Sweet et al.\ (2018) found a discrepant exponent ($\alpha \approx 1$) for disk-dominated galaxies.
We discuss another reason in the Appendix. 

The third conclusion apparent from Figure~1 is that our $j_{\star}$--$M_{\star}$ relations for disk-dominated and bulge-dominated galaxies are roughly, but not exactly, parallel to each other. 
We quantify this impression by fitting Equation~(1) to the galaxies in our sample with $\beta_\star \geq 0.8$, i.e., essentially pure bulges (within the uncertainties in $\beta_\star$), finding $\alpha = 0.83\pm0.16$ and $\log j_0 = 2.20\pm0.12$.
Evidently, the $j_{\star}$--$M_{\star}$ relation for bulge-dominated galaxies is slightly steeper than the one for disk-dominated galaxies, although the two exponents are consistent with each other at the $\sim 1.5 \sigma$ level, while the normalizations differ by a factor of $7\pm2$.
The power-law fit for bulge-dominated galaxies presented here differs slightly from the one given in Paper~2, because the newer fit is based on a subsample of galaxies defined by a strict limit on bulge fraction, while the older one was based on a subsample comprised of all elliptical galaxies irrespective of their bulge fractions.

Unfortunately, we cannot compare our $j_{\star}$--$M_{\star}$ scaling relation for bulge-dominated galaxies with other determinations, because none of the other samples includes enough high-$\beta_\star$ galaxies.
We note that Cortese et al.\ (2016) found roughly parallel $j_{\star}$--$M_{\star}$ relations for galaxies of different morphologies, ranging from early-type spirals to ellipticals, based on absorption-line kinematics derived from SAMI\footnote{SAMI is an acronym for Sydney-AAO (Anglo-Australian Observatory) Multi-object Integral field spectrograph} integral-field spectroscopy, in qualitative agreement with our results.
However, the galaxies in the Cortese et al.\ (2016) sample lack disk--bulge decompositions and kinematic data that reach beyond $\sim 1 R_{\rm e}$, thus precluding a quantitative comparison with our results.

\begin{figure*}
\includegraphics[width=\textwidth]{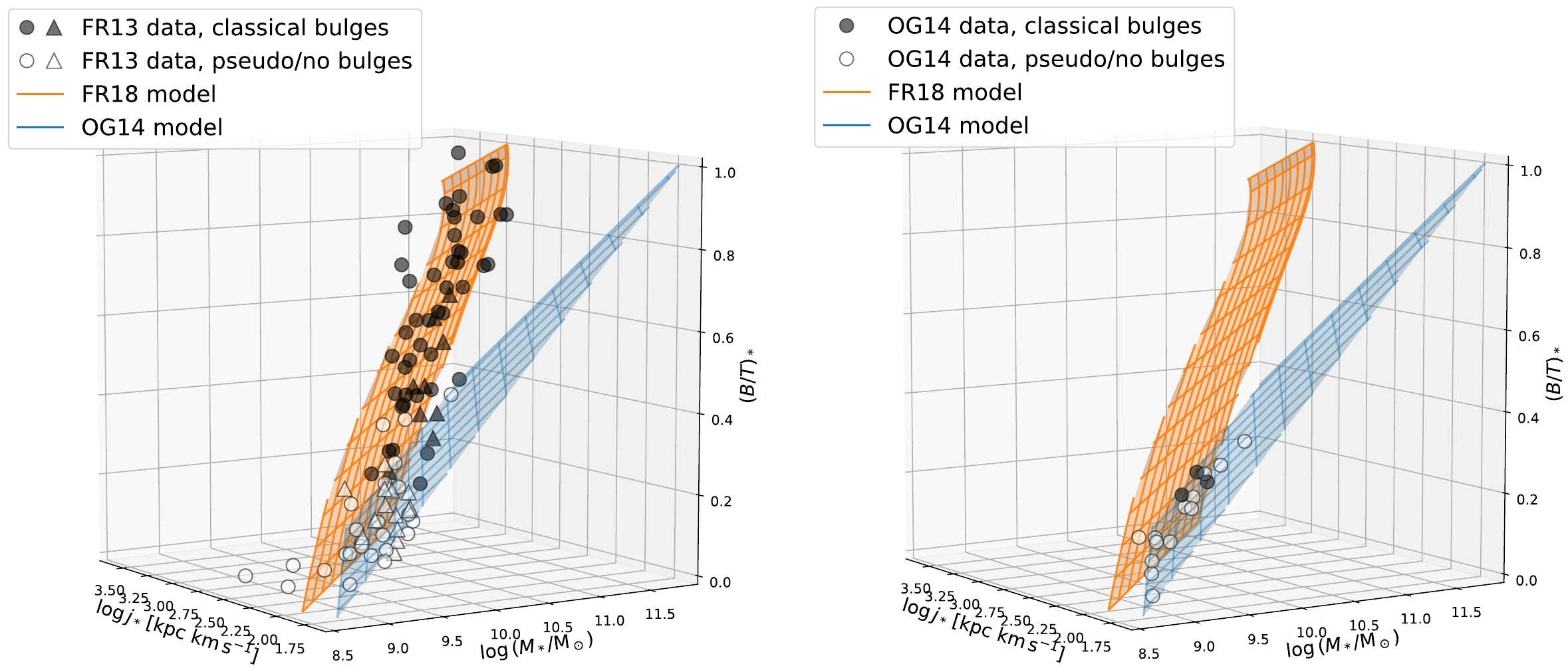}
\caption{
Stellar bulge fraction, $\beta_\star \equiv (B/T)_{\star}$, plotted against stellar specific angular momentum $j_{\star}$ and stellar mass $M_{\star}$ from our work (left panel) and from Obreschkow \& Glazebrook (2014) (right panel).  
Filled symbols represent galaxies with classical bulges, and open symbols those with pseudo bulges or no bulges.  
For circles, the bulge classifications are more certain; for triangles, they are less certain.
In both panels, the orange surface represents the relation for independent disks and bulges derived in the present study, while the blue plane represents the linear regression derived by Obreschkow \& Glazebrook (2014).  
}
\end{figure*}

\section{3D Relations Between  \MakeLowercase{\it J}$_{\star}$,  $M_{\star}$, {\rm and} $\beta_{\star}$}

We now examine the distribution of galaxies in the 3D space of specific angular momentum $j_{\star}$, mass $M_{\star}$, and bulge fraction $\beta_\star$.
The $j_{\star}$--$M_{\star}$ scaling relations discussed in the previous section are simply the projections of this distribution along the $\beta_\star$ axis.
We focus mainly on our own dataset, because it is the only one with full coverage in $\beta_\star$.
However, we compare our results directly with those of Obreschkow \& Glazebrook (2014) because they have made testable claims about the distribution of galaxies in ($\log j_{\star}$, $\log M_{\star}$, $\beta_\star$) space.
In the following, we disregard the Posti et al.\ (2018a) dataset because of its narrow coverage in $\beta_\star$ and the Sweet et al.\ (2018) dataset for the reasons discussed in Section 2 and the Appendix.

The left and right panels of Figure~2 show the distribution of galaxies in ($\log j_{\star}$, $\log M_{\star}$, $\beta_\star$) space for our sample and for the Obreschkow \& Glazebrook (2014) sample, respectively.
Galaxies with classical and pseudo bulges are distinguished in this diagram by filled and open symbols, respectively. 
Our sample has a mixture of classical and pseudo bulges, while the Obreschkow \& Glazebrook (2014) sample is dominated by pseudo bulges. 
The galaxies in our sample lie on or near the curved orange surface, while those in the Obreschkow \& Glazebrook (2014) sample lie on or near the blue plane.
The orange surface and blue plane, which are replicated in both panels of Figure~2, differ substantially at high bulge fraction, but converge toward each other at low bulge fraction.
We elaborate on these similarities and differences in the remainder of this section.

The orange surface in Figure~2 is based on the following simple model inspired by the similar $j_{\star}$--$M_{\star}$ scaling relations of galaxies with different $\beta_\star$ shown in Figure~1.
In this model, normal galaxies, in a first approximation, consist of a linear superposition of disks and bulges that follow separate scaling relations of the form $j_{\star{\rm d}}  =  j_{0{\rm d}} (M_{{\star}{\rm d}}/M_0)^{\alpha}$ and $j_{\star{\rm b}}  =  j_{0{\rm b}} (M_{{\star}{\rm b}}/M_0)^{\alpha}$, respectively (with $\log (M_0/M_\odot) = 10.5$ and all specific angular momenta expressed in units of kpc~km~s$^{-1}$, as before).
Then the total values of specific angular momentum $j_{\star}$ and mass $M_{\star}$ for composite galaxies of any bulge fraction $\beta_\star$ are related by
\begin{equation}
\begin{split}
j_{\star}  & =   ( j_{\star{\rm d}} M_{{\star}{\rm d}}  +  j_{\star{\rm b}} M_{{\star}{\rm b}} ) / ( M_{{\star}{\rm d}}  +  M_{{\star}{\rm b}} )  \\
& =    j_0(\beta_\star) (M_{\star}/M_0)^{\alpha},
\end{split}
\end{equation}
\begin{equation}
j_0(\beta_\star)  =  j_{0{\rm d}} (1 - \beta_\star)^{\alpha + 1}  +  j_{0{\rm b}} \beta_\star^{\alpha + 1}.
\end{equation}

As expected, this $j_{\star}$--$M_{\star}$ relation has the same exponent $\alpha$ for all values of $\beta_\star$, corresponding to parallel lines in the $\log j_{\star}$--$\log M_{\star}$ diagram.
The dependence of $\log j_{\star}$ on $\beta_\star$ at fixed $\log M_{\star}$, however, is non-linear.
Thus, in the 3D space of $\log j_{\star}$, $\log M_{\star}$, and $\beta_\star$, galaxies will lie on or near a curved 2D surface given by Equations (2) and (3) if they obey this simple model.
It is easy to generalize this model to one in which the scaling relations for disks and bulges have different exponents, $\alpha_{\rm d}$ and $\alpha_{\rm b}$.
We have not done this because it only complicates the analysis, without a commensurate gain in accuracy or insight, and because, with currently available data, $\alpha_{\rm d}$ and $\alpha_{\rm b}$ are statistically equal at the $\sim 1.5\sigma$ level according to our 2D fits in Section~2.2.

The orange surface in Figure~2 is our 3D fit of this simple model to our full dataset ($8.9 \leq \log ( M_{\star} /M_{\odot} ) \leq 11.8$). 
In particular, we derive the best-fit values and $1\sigma$ errors of the parameters $\alpha$, $j_{0{\rm d}}$, and $j_{0{\rm b}}$ in Equations (2) and (3) by minimizing the trivariate $\chi^2$ with the observations in the form ($\log j_{\star}$, $\log M_{\star}$, $\beta_\star$) and our estimates of the typical errors from Section~2.1.
The results are $\alpha = 0.67 \pm 0.07$, $\log j_{0{\rm d}} = 3.17 \pm 0.03$, $\log j_{0{\rm b}} = 2.25 \pm 0.14$, and $\chi_{\rm red}^2 = 1.0$.
The values of $\alpha$ and $j_{0{\rm b}}$ from this 3D fit are statistically the same as those from the 2D fits in Section~2.2 (within $1 \sigma$), while the value of $\log j_{0{\rm d}}$ is $0.10 \pm 0.04$ higher.
As expected, the single exponent $\alpha$ for all galaxies in the 3D fit lies between the separate values of $\alpha$ for disk-dominated and bulge-dominated galaxies in the 2D fits.  
The normalizations for disks and bulges ($j_{0{\rm d}}$ and $j_{0{\rm b}}$) in the 3D fit differ by a factor of $8\pm2$, slightly higher than in the 2D fits. 

Two other features of this 3D fit are noteworthy.  
First, because the fit has $\chi_{\rm red}^2 = 1.0$, all the observed scatter can be accounted for by the estimated (total) errors, $\varepsilon(\log j_\star)$, $\varepsilon(\log M_\star)$, and $\varepsilon(\beta_\star)$, with no need to invoke any intrinsic scatter.
However, because these errors are only approximate, we cannot rule out even a fairly large intrinsic scatter, similar to the errors themselves.  
Second, the agreement between the model and our dataset is not simply a consequence of the fact that, in both cases, the total specific angular momentum and mass of galaxies ($j_\star$, $M_\star$) represent sums over the contributions from disks ($j_{{\star}{\rm d}}$, $M_{{\star}{\rm d}}$) and bulges ($j_{{\star}{\rm b}}$, $M_{{\star}{\rm b}}$).
In the model, disks and bulges are assumed to follow separate scaling relations, with $j_{\star{\rm d}}$ determined only by $M_{\star{\rm d}}$, and $j_{\star{\rm b}}$ only by $M_{\star{\rm b}}$, whereas no such assumption was made in the empirical estimates of $j_\star$ and $M_\star$.
Moreover, for lenticular and elliptical galaxies, we estimated $j_\star$ and $M_\star$ directly from the overall surface brightness and stellar rotation profiles, without distinguishing the contributions from disks and bulges.

\begin{figure*}
\includegraphics[width=\textwidth]{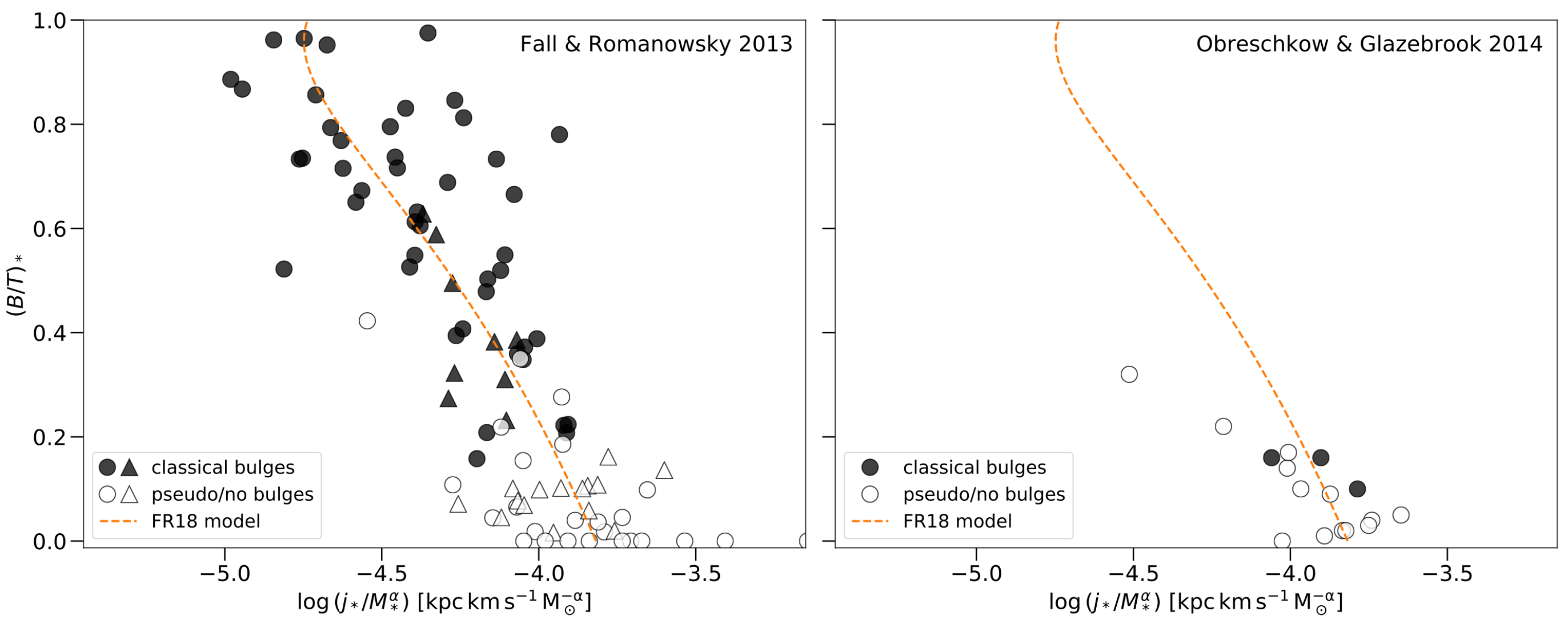}
\caption{
Stellar bulge fraction, $\beta_\star \equiv (B/T)_{\star}$, plotted against $j_{\star}/M_{\star}^{\alpha}$ with $\alpha = 0.67$ from our work (left panel) and from Obreschkow \& Glazebrook (2014) (right panel).  
Filled symbols represent galaxies with classical bulges, and open symbols those with pseudo bulges or no bulges.  
For circles, the bulge classifications are more certain; for triangles, they are less certain.  
In both panels, the orange dashed line represents the relation for independent disks and bulges derived in the present study. 
Note that galaxies with pseudo bulges cluster at the lower end of this relation.
}
\end{figure*}

The blue plane in Figure~2 is the 3D linear regression by Obreschkow \& Glazebrook (2014) to their dataset.
They derived best-fit parameters ($k_1$, $k_2$, $k_3$) of the plane $\beta_\star = k_1 \log M_\star + k_2 \log j_\star + k_3$ by minimizing the trivariate $\chi^2$ with respect to the observations in the form ($\log j_{\star}$, $\log M_{\star}$, $\beta_\star$) and using their estimates of the typical errors.
They then reexpressed these results in the form $j_{\star}  =  j_0(\beta_\star) (M_{\star}/M_0)^{\alpha}$, analogous to our Equation (2), but with $j_0(\beta_\star) = k \exp(-g\beta_\star) \times 1000$~kpc~km~s$^{-1}$, in place of our Equation (3), and the parameter values $\alpha = 0.94 \pm 0.07$, $k = 0.89 \pm 0.11$, $g = 7.03 \pm 1.35$ (with $\log (M_0/M_\odot) = 10.0$).
It is worth noting that this fit is not motivated by any underlying physical model; it is simply a convenient representation of the data.
Moreover, the robustness of the fit is questionable, given that it is based on a small sample of galaxies ($N = 16$) with small bulge fractions ($\beta_\star \leq 0.3$). 

It is clear at a glance that the blue plane derived from the Obreschkow \& Glazebrook (2014) dataset is not an acceptable fit to our dataset, especially for large bulge fractions.
However, it is less obvious whether the orange surface derived from our dataset provides a good or bad fit to their dataset.
This depends critically on the adopted errors in the $\chi^2$ calculation.
If we adopt the typical errors quoted by Obreschkow \& Glazebrook (2014) of 0.02, 0.06, 0.02, respectively, for $\log j_{\star}$, $\log M_{\star}$, $\beta_\star$, the fit is rejected by a wide margin. 
However, as discussed in Section 2 and the Appendix, these (partial) errors are unrealistically small. 
In particular, they are much smaller than the dispersions between independent estimates of these quantities by different authors. 
If instead, we adopt our estimates of the typical (total) errors from Section~2.1, we obtain $\chi_{\rm red}^2 = 0.9$, indicating  an acceptable fit of the orange surface to the Obreschkow \& Glazebrook (2014) dataset.

In Figure~3, we plot $\beta_\star$ against $\log (j_{\star} / M_{\star}^{\alpha})$ with $\alpha = 0.67$ for both our dataset and the Obreschkow \& Glazebrook (2014) dataset.
This 2D projection of the 3D space of $j_{\star}$, $M_{\star}$, and $\beta_\star$ effectively removes the primary dependence of $j_{\star}$ on $M_{\star}$ from the scaling relation, thus highlighting its secondary dependence on $\beta_\star$.
The dashed curves in Figure~3, computed from Equations (2) and (3), are the projection of the orange surface in this direction.
This representation shows even more clearly than Figure~2 that both datasets are consistent, within the scatter, with our simple model based on independent disks and bulges. 
Given the low $\chi_{\rm red}^2$, even the lone outlier in the Obreschkow \& Glazebrook (2014) dataset is consistent with this model.

In Figures~2 and~3, we distinguish classical bulges from pseudo bulges by filled and open symbols, respectively.
The bulge types for galaxies in our sample and the references from which they were taken, if available, are listed in Table~1. 
For 23 of the spirals, there are definite bulge classifications in the literature.
Another 10 have no bulges, according to our adopted decompositions.
The remaining 24 spirals either have no available bulge classifications, or have ambiguous types in the literature.
We tentatively classify the bulges of these galaxies as ``pseudo?" if they have $\beta_\star < 0.2$ and ``classical?" if they have $\beta_\star > 0.2$, as suggested by the $\beta_\star$-distributions of galaxies with definite bulge types.
In Figures~2 and~3, galaxies with more certain bulge types are indicated by circles, while those with less certain bulge types are indicated by triangles.
 
It is worth noting here that the definition, observational signatures, and physical origins of pseudo bulges are not universally agreed upon (see Kormendy \& Kennicutt 2004 for a review).
For some authors, pseudo bulges are flattened (disk-like) structures in the inner parts of galaxies for which the surface brightness profile exceeds that of a fitted exponential disk.
For others, they are bulges for which the fitted surface brightness profile has a low S\'ersic index (typically, $n < 2$), irrespective of whether they have flat or round 3D shapes.
For still others, pseudo bulges are whatever structures may result from the ``secular'' re-arrangement of material in the inner parts of galactic disks by bars and/or oval distortions, including heating in the vertical direction. 
Evidently, there is no consensus on even such a basic property of pseudo bulges as their 3D shapes. 
These ambiguities complicate any attempts to discern whether classical and pseudo bulges follow the same or different $j_{\star}$--$M_{\star}$ scaling relations and to interpret such results astrophysically. 

With these caveats in mind, we note from Figure~2, and especially Figure~3, the strong tendency for galaxies with classical and pseudo bulges to be segregated from each other above and below $\beta_\star \sim 0.2$.
At the same time, they appear to span {\it different parts} of the {\it same surface} in ($\log j_{\star}$, $\log M_{\star}$, $\beta_\star$) space.
In particular, they are both consistent, within the scatter, with our simple model based on independent disks and bulges, with $\chi_{\rm red}^2 = 0.7$ for classical bulges and $\chi_{\rm red}^2 = 1.2$ for pseudo bulges. 
In any case, there is no statistically significant indication, from either our dataset or the Obreschkow \& Glazebrook (2014) dataset, that galaxies with classical and pseudo bulges lie on fundamentally different 2D surfaces in the 3D space of $j_{\star}$, $M_{\star}$, and $\beta_\star$.  
The scatter in the observations is large enough, however, that we cannot rule out subtle differences in these distributions.

\section{Summary and Discussion}

The main conclusion of this paper is that the observed values of specific angular momentum $j_{\star}$, mass $M_{\star}$, and bulge fraction $\beta_{\star}$ of the stellar parts of most normal galaxies are consistent, in a first approximation, with a simple model based on a linear superposition of independent disks and bulges.
The disks and bulges in this model follow scaling relations of the form $j_{\star{\rm d}} \propto M_{\star{\rm d}}^{\alpha}$ and $j_{\star{\rm b}} \propto M_{\star{\rm b}}^{\alpha}$ with $\alpha = 0.67 \pm 0.07$ but offset from each other by a factor of $8 \pm 2$ over the mass range $8.9 \leq \log (M_{\star}/M_{\odot}) \leq 11.8$.
Separate fits for disks and bulges alone give $\alpha = 0.58 \pm 0.10$ and  $\alpha = 0.83 \pm 0.16$, respectively.
This simple model correctly predicts that galaxies will lie on or near a curved 2D surface specified by Equations (2) and (3) in the 3D space of $\log j_{\star}$, $\log M_{\star}$, and $\beta_\star$.
These results reinforce and extend our earlier suggestion that the distribution of galaxies with different $\beta_{\star}$ in the $j_{\star}$--$M_{\star}$ diagram constitutes an objective, physically motivated alternative to subjective classification schemes such as the Hubble sequence.

For disk-dominated galaxies in the mass range considered here, the $j_{\star}$--$M_{\star}$ scaling relation is now quite secure, as shown in Figure~1 by the excellent agreement between our determination (from Paper~2) and those of Obreschkow \& Glazebrook (2014) and Posti et al.\ (2018a).
Two factors contribute to the robustness of this scaling relation.
First, no special efforts are required to obtain photometric and kinematic data that extend to large enough radii (in units of $R_{\rm e}$) to estimate reliably the disk contributions $j_{{\star}{\rm d}}$ and $M_{{\star}{\rm d}}$ to the total values $j_{\star}$ and $M_{\star}$, which usually turn out to be close to those for ideal disks with exponential surface density profiles and flat rotation curves.
Second, any uncertainties in the bulge contributions $j_{{\star}{\rm b}}$ and $M_{{\star}{\rm b}}$, even when substantial, have only a minor impact on the total values $j_{\star}$ and $M_{\star}$. 

For most of the giant galaxies studied here, cold gas (\hi\ and H$_2$) makes a relatively small contribution to their specific angular momentum and mass, and the stellar $j_{\star}$--$M_{\star}$ scaling relation is a good proxy for the baryonic $j_{\rm bary}$--$M_{\rm bary}$ scaling relation (Obreschkow \& Glazebrook 2014).\footnote
{The baryonic scaling relations mentioned here include only stars and cold gas, not the warm and hot diffuse gas in galactic halos, which might actually dominate the total $j_{\rm bary}$ and $M_{\rm bary}$ budgets of some galaxies.}
This is no longer true, however, for gas-rich dwarf galaxies, which contain more specific angular momentum and mass in cold gas than in stars.
Several recent studies have extended the $j_{\rm bary}$--$M_{\rm bary}$ scaling relation down into the mass range $7 \lea \log (M_{\rm bary}/M_{\odot}) \lea 9$, with somewhat confusing claims about whether it lies above or matches onto the extrapolated $j_{\rm bary}$--$M_{\rm bary}$ and $j_{\star}$--$M_{\star}$ scaling relations from higher masses (Butler et al.\ 2017; Chowdhury \& Chengalur 2017; Elson 2017; Kurapati et al.\ 2018).
Continuing and refining this work is important, because it has the potential to place constraints on the mass dependence of the retained or sampled fraction of specific angular momentum in galaxies $f_j$ (see below).

For bulge-dominated galaxies, the $j_{\star}$--$M_{\star}$ scaling relation is based almost entirely on our work.
In this case, the main challenge is obtaining kinematic data that extend to large enough radii (in units of $R_{\rm e}$) that the estimates of $j_{{\star}{\rm b}}$ have converged.  
This is important because the stellar rotation profiles of bulge-dominated galaxies, unlike the H$\alpha$ and HI rotation curves of disk-dominated galaxies, exhibit a great variety of behaviors; some are flat, while others rise or fall.
All of our estimates of $j_{{\star}{\rm b}}$ are based on kinematic data that extend to $\sim 2R_{\rm e}$ and some to much larger radii, thus capturing as much angular momentum with as little extrapolation as possible. 
Nevertheless, additional studies of the $j_{\star}$--$M_{\star}$ scaling relation for spheroid-dominated galaxies, based on 2D kinematic data that reach even larger radii (for example, $R > 5 R_{\rm e}$), would certainly be desirable. 

For intermediate-type galaxies, the main challenge to deriving the $j_{\star}$--$M_{\star}$ scaling relation is in disentangling the contributions to $j_{\star}$ and $M_{\star}$ from the superposed disks and bulges.
The specific angular momenta of bulges in such galaxies have been approximated in three different ways, by assuming that their rotation velocity is either (1) zero, (2) the same as the rotation velocity of their associated disks, or (3) the same as the mean rotation velocity for bulges of the same velocity dispersion and ellipticity.
Method (1) and (2) clearly lead to systematic under- and overestimates of $j_{{\star}{\rm b}}$, respectively, while method (3), the one we have adopted, contributes some scatter but little if any bias to the $j_{\star}$--$M_{\star}$ scaling relation.
More accurate results will require careful modeling of extensive 2D photometric and kinematic data to disentangle the velocity fields and hence the specific angular momenta of superposed disks and bulges (as in the recent work of Rizzo et al.\ (2018) on lenticular galaxies).

The bulge fraction $\beta_\star$ is inherently uncertain because it depends on the adopted method for decomposing galaxies into disks and bulges, either by pre-specifying their 3D shapes (flat versus round) or by pre-specifying their surface brightness profiles (exponential versus S\'ersic).
These two methods generally give similar values of $\beta_\star$ for bulge-dominated galaxies (elliptical, lenticulars, and early-type spirals), but they can give substantially different values of $\beta_\star$ for disk-dominated galaxies (late-type spirals).
A related complication is the lack of consensus on the definition of pseudo bulges, including whether they must always be flat (like disks) or may sometimes be round (like spheroids).
This ambiguity adds substantially to the uncertainty in estimates of $\beta_\star$ for disk-dominated galaxies, where pseudo bulges are much more common than classical bulges.   

We find no statistically significant indication that galaxies with pseudo bulges and classical bulges follow different relations in the space of $\log j_{\star}$, $\log M_{\star}$, and $\beta_\star$.
This does not mean that both types of galaxies follow exactly the same relation, of course, merely that any differences must be small enough to hide within the scatter.
Obreschkow \& Glazebrook (2014) found a different relation (with $\alpha \approx 1$) from a small sample of spiral galaxies with a preponderance of pseudo bulges (13/16) covering a narrow range in $\beta_\star$.
This result, however, is based on adopted (partial) errors in $j_\star$, $M_\star$, and $\beta_\star$ that neglect the inherent uncertainties mentioned above and are therefore unrealistically small. 
As we have shown here, the statistical significance of the Obreschkow \& Glazebrook (2014) relation disappears when we adopt more realistic (total) errors in these quantities. 
Sweet et al.\ (2018) also found a different relation (again with $\alpha \approx 1$), based on a dataset with large systematic errors in $j_{\star}$.

Finally, we offer a few remarks on the astrophysical implications of our results, following the precepts of Paper~1.
Comparing the scaling relation for the stellar components of galaxies in the form $j_\star = j_0 (M_\star / M_0)^{\alpha}$ with that for dark-matter halos in the standard $\Lambda$CDM cosmology, we derive the relation $f_j / f_M^{2/3} = 6.8 (j_0/10^3 \,{\rm kpc\,km\,s^{-1}})(M_\star / M_0)^{\alpha -2/3}$ between the fractions of specific angular momentum and mass in stars relative to dark matter, $f_j \equiv j_\star / j_{\rm halo}$ and $f_M \equiv M_\star / M_{\rm halo}$ (with $M_0 = 10^{10.5} M_\odot$ again).
With the exponent $\alpha$ and normalizations $j_0$ from our 3D fit to Equations~(2) and~(3), this relation becomes $f_j / f_M^{2/3} = 10.0 \pm 0.6$ for disks and $f_j / f_M^{2/3} = 1.2 \pm 0.4$ for bulges.
Then, with the separate relations between $f_M$ and $M_\star$ for late-type and early-type galaxies from Dutton et al.\ (2010), we obtain $f_j \approx 1.0$ for disks and $f_j \approx 0.1$ for bulges at $M_\star \sim 10^{10.5} M_\odot$ and only mild variations over the range $10^{9.5} M_\odot \lea M_\star \lea 10^{11.5} M_\odot$.
These estimates of $f_j$ differ slightly from the ones derived in Paper~2 for the same dataset because the new 3D and old 2D fits return slightly different values of $j_0$.    

The relations $f_j / f_M^{2/3} = {\rm constant}$ derived above imply that $f_j$ and $f_M$ must have qualitatively similar dependences on $M_\star$, namely a broad peak near $M_\star \sim 10^{10.5} M_\odot$, a shallow decline to lower $M_\star$, and a somewhat steeper decline to higher $M_\star$.
This is why we find only mild variations in $f_j$.
Recent analyses by other authors also indicate $f_j \approx {\rm constant}$ near $M_\star \sim 10^{10.5} M_\odot$ (see Fig.\ 12 of Lapi et al.\ 2018b and Fig.\ 3 of Posti et al.\ 2018a).
Over much wider mass ranges, the deviations from a constant $f_j$ may become more pronounced.  
The model of disk formation preferred by Posti et al.\ (2018a) has $f_j \propto f_M^s \propto M_{\star}^{\gamma}$ with $\gamma = s (2 - 3\alpha)/(2 - 3s)$, which, when fitted to their full dataset ($10^{7.0} M_\odot \leq M_\star \leq 10^{11.3} M_\odot$), gives $\alpha =0.59 \pm 0.02$, $s = 0.4 \pm 0.1$, and thus $\gamma = 0.12 \pm 0.03$.
However, even this weak dependence of $f_j$ on mass could be erased if the $j_{\rm bary}$--$M_{\rm bary}$ relation for gas-rich dwarf galaxies turns out to be shallower than the $j_\star$--$M_\star$ relation by only $\Delta\alpha \approx 0.15$ (again, see Fig.\ 3 of Posti et al.\ 2018a).
This is why it is important to refine estimates of the baryonic relation at low masses.   
  
The fractions $f_j$ and $f_M$ for disks and bulges and the corresponding $j$--$M$ scaling relations (stellar and baryonic) are potentially determined by a large number of astrophysical processes.
These include tidal torques, dynamical friction of baryonic structures within dark-matter halos, shocks and radiative cooling in the interstellar and circumgalactic media, star formation and its associated feedback, inflow, outflow, and recycling of gas, merging of gas clumps and dwarf galaxies, and tidal stripping of the outer parts of halos and their circumgalactic media by neighboring halos.  
We reviewed these processes and their potential impact on the $j$--$M$ scaling relations for disks and bulges at some length in Paper~1.
Here, we note only the growing interest in biased-collapse models in which the fractions $f_j$ and $f_M$ are determined by the hypotheses that the baryons and dark matter in protogalaxies start with similar distributions of specific angular momentum and mass and that, at any given time, only the baryons within some critical radius are able to collapse and form the visible parts of galaxies.
Analytical models of this type and their implications for the $j$--$M$ scaling relations are explored in several recent papers (Shi et al.\ 2017; Lapi et al.\ 2018a; Posti et al.\ 2018a, 2018b; see also Paper~1 and references therein).
 
In the past few years, hydrodynamical simulations of forming galaxies have succeeded in reproducing, at least approximately, the observed $j$--$M$ scaling relations (Genel et al.\ 2015; Pedrosa \& Tissera 2015; Teklu et al.\ 2015; Agertz \& Kravtsov 2016; Zavala et al.\ 2016; DeFelippis et al.\ 2017; Grand et al.\ 2017; Lagos et al.\ 2017; Soko{\l}owska et al.\ 2017; Stevens et al.\ 2017; El-Badry et al.\ 2018; Lagos et al.\ 2018; Obreja et al.\ 2018).
One of the main lessons from these simulations is that feedback in an essential ingredient to match the observed relations for both disk-dominated and spheroid-dominated galaxies. 
Without feedback, the simulations suffer from the well-known overcooling and angular momentum problems and fail to produce the full range of galactic morphologies.
Another important ingredient is merging, which appears to explain, at least partially, the slow rotation of spheroids relative to disks.
 
Despite the success of recent analytical models and hydrodynamical simulations, we do not yet have definitive answers to some important theoretical questions about the $j$--$M$ scaling relations, such as the following. 
Given the potential complexity of galaxy formation, why are the observed $j$--$M$ relations so simple? 
In particular, why are the specific angular momentum and mass fractions $f_j$ and $f_M$ so closely linked that they result in power-law $j$--$M$ relations over 3--4 decades in mass (at least for disks)?
Why do the disks of massive galaxies have nearly the same specific angular momentum as their dark-matter halos ($f_j \sim 1.0$) and why do their bulges have much less ($f_j \sim 0.1$)?
Answering these questions will require a better understanding of how much the specific angular momentum of mass elements inside forming galaxies is redistributed and which physical mechanisms are most responsible for this redistribution. 
This is a promising direction for future analysis of hydrodynamical simulations (as already begun by DeFelippis et al.\ 2017).

\acknowledgements

We thank Kenneth Freeman and John Kormendy for guiding us through the mysteries of pseudo bulges.
This research was supported in part by the National Science Foundation through grants AST16-16710 and PHY17-48958.
AJR is a Research Corporation for Science Advancement Cottrell Scholar.


\bibliographystyle{aasjournal}

\clearpage
\newpage

\startlongtable
\begin{deluxetable*}{l c c c c c l }
\tablecaption{Specific Angular Momenta, Masses, and Bulge Fractions of Sample Galaxies\label{tab:bulges}}
\tablehead{
\colhead{Name} & \colhead{Galaxy type} & \colhead{$\log_{10} M_\star$} &\colhead{$\log_{10} j_\star$} & \colhead{$\beta_\star$} & \colhead{Bulge type} & \colhead{Reference}  \\
& &  [$M_\odot$] & [kpc~km~s$^{-1}$] & & & \\
}
\colnumbers
\vspace{-0.1cm}
\startdata
\vspace{-0.1cm}
NGC 224 & Sb & 10.91 & 3.34 & 0.22 & classical & KK04, FD11 \\ 
\vspace{-0.1cm}
NGC 247 & Sd & 9.47 & 2.90 & 0.00 & none & FD11  \\ 
\vspace{-0.1cm}
NGC 300 & Sd & 8.95 & 2.42 & 0.00 & none & FD11 \\ 
\vspace{-0.1cm}
NGC 701 & Sc & 10.20 & 2.74 & 0.00 & none & --  \\
\vspace{-0.1cm}
NGC 753 & Sbc & 11.00 & 3.27 & 0.07 & pseudo? & --  \\ 
\vspace{-0.1cm}
NGC 801 & Sc & 11.30 & 3.60 & 0.28 & pseudo & KB11 \\ 
\vspace{-0.1cm}
NGC 821 & E6 & 10.94 & 2.53 & 0.73 & classical & dS+04  \\ 
\vspace{-0.1cm}
NGC 1023 & S0 & 10.92 & 3.19 & 0.67 & classical & K+11, F+12, C+13 \\ 
\vspace{-0.1cm}
NGC 1024 & Sab & 11.21 & 3.32 & 0.38 & classical? & --  \\
\vspace{-0.1cm}
NGC 1087 & Sc & 10.25 & 2.91 & 0.00 & none & -- \\
\vspace{-0.1cm}
NGC 1316 & S0 & 11.85 & 3.64 & 0.81 & classical & dS+04 \\ 
\vspace{-0.1cm}
NGC 1325 & Sbc & 10.35 & 3.16 & 0.05 & pseudo & FD08 \\ 
\vspace{-0.1cm}
NGC 1339 & E3 & 10.44 & 2.81 & 0.73 & classical & --  \\
\vspace{-0.1cm}
NGC 1344 & E4 & 11.05 & 2.59 & 0.73 & classical & --  \\
\vspace{-0.1cm}
NGC 1353 & Sbc & 10.65 & 3.04 & 0.15 & pseudo & KK04, FD08 \\ 
\vspace{-0.1cm}
NGC 1357 & Sab & 10.92 & 3.22 & 0.37 & classical & GH17 \\ 
\vspace{-0.1cm}
NGC 1373 & E2 & 9.71 & 1.62 & 0.96 & classical & --  \\
\vspace{-0.1cm}
NGC 1379 & E0 & 10.60 & 2.08 & 0.89 & classical & --  \\
\vspace{-0.1cm}
NGC 1380 & S0 & 11.27 & 3.45 & 0.35 & classical & --  \\
\vspace{-0.1cm}
NGC 1381 & S0 & 10.68 & 3.10 & 0.39 & classical & --  \\
\vspace{-0.1cm}
NGC 1400 & S0 & 10.99 & 2.37 & 0.87 & classical & --  \\
\vspace{-0.1cm}
NGC 1404 & E1 & 11.18 & 3.02 & 0.83 & classical & --  \\
\vspace{-0.1cm}
NGC 1407 & E0 & 11.58 & 3.03 & 0.95 & classical & --  \\
\vspace{-0.1cm}
NGC 1417 & Sb  & 11.05 & 3.51 & 0.11 & pseudo? & -- \\
\vspace{-0.1cm}
NGC 1421 & Sbc & 10.42 & 3.33 & 0.14 & pseudo? & -- \\
\vspace{-0.1cm}
NGC 1620 & Sbc & 11.08 & 3.51 & 0.10 & pseudo? & --  \\
\vspace{-0.1cm}
NGC 2310 & S0  & 10.25 & 2.91 & 0.21 & classical & -- \\
\vspace{-0.1cm}
NGC 2403 & Scd &  9.58 & 2.70 & 0.00 & none & --  \\ 
\vspace{-0.1cm}
NGC 2577 & S0 & 10.75 & 3.08 & 0.35 & classical & -- \\
\vspace{-0.1cm}
NGC 2590 & Sbc & 11.17 & 3.33 & 0.31 & classical? &  \\
\vspace{-0.1cm}
NGC 2592 & E2 & 10.68 & 3.00 & 0.55 & classical & MA+18  \\ 
\vspace{-0.1cm}
NGC 2608 & Sb & 10.44 & 2.86 & 0.10 & pseudo? & --  \\ 
\vspace{-0.1cm}
NGC 2639 & Sa & 11.24 & 3.11& 0.63 & classical? & --  \\
\vspace{-0.1cm}
NGC 2699 & E1 & 10.47 & 2.59 & 0.61 & classical & --  \\
\vspace{-0.1cm}
NGC 2708 & Sb & 10.63 & 3.14 & 0.10 & pseudo? & --  \\
\vspace{-0.1cm}
NGC 2715 & Sc & 10.46 & 3.20 & 0.02 & pseudo? & --  \\
\vspace{-0.1cm}
NGC 2742 & Sc & 10.33 & 3.08 & 0.02 & pseudo & M04 \\ 
\vspace{-0.1cm}
NGC 2768 & S0 & 11.28 & 3.57 & 0.78 & classical & C+13  \\ 
\vspace{-0.1cm}
NGC 2775 & Sab & 11.22 & 3.30 & 0.21 & classical & W+09, F+12 \\ 
\vspace{-0.1cm}
NGC 2778 & E2 & 10.18 & 2.66 & 0.52 & classical & --  \\
\vspace{-0.1cm}
NGC 2815 & Sb  & 11.08 & 3.30 & 0.39 & classical? & -- \\
\vspace{-0.1cm}
NGC 2841 & Sb & 10.99 & 3.05 & 0.39 & classical & F+12  \\ 
\vspace{-0.1cm}
NGC 2844 & Sa & 10.16 & 2.66 & 0.23 & classical? & --  \\
\vspace{-0.1cm}
NGC 2903 & Sbc & 10.49 & 3.00 & 0.00 & none  & -- \\ 
\vspace{-0.1cm}
NGC 2998 & Sc & 10.79 & 3.37 & 0.04 & pseudo & KB11 \\
\vspace{-0.1cm}
NGC 3031 & Sab & 10.85 & 3.02 & 0.16 & classical & KK04, F+12 \\ 
\vspace{-0.1cm}
NGC 3067 & Sab & 10.35 & 2.77 & 0.05 & pseudo? & --  \\
\vspace{-0.1cm}
NGC 3115 & S0 & 10.98 & 3.14 & 0.50 & classical & FD08, K+11, C+13 \\ 
\vspace{-0.1cm}
NGC 3156 & S0 & 10.08 & 2.47 & 0.41 & classical & --  \\
\vspace{-0.1cm}
NGC 3198 & Sc & 10.11 & 3.02 & 0.00 & none & --  \\ 
\vspace{-0.1cm}
NGC 3200 & Sc &  11.13 & 3.62 & 0.16 & pseudo? & -- \\
\vspace{-0.1cm}
NGC 3203 & S0 & 10.84 & 3.30 & 0.22 & classical & --  \\
\vspace{-0.1cm}
NGC 3377 & E5 & 10.42 & 2.53 & 0.53 & classical & --  \\
\vspace{-0.1cm}
NGC 3379 & E2 & 10.88 & 2.53 & 0.86 & classical & -- \\ 
\vspace{-0.1cm}
NGC 3593 & S0/a  & 9.78 & 2.25 & 0.07 & pseudo? & KB11, S+18 \\ 
\vspace{-0.1cm}
NGC 3605 & E3  & 10.01 & 2.28 & 0.63 & classical & -- \\
\vspace{-0.1cm}
NGC 3898 & Sab & 10.98 & 3.01 & 0.69 & classical & F+12  \\ 
\vspace{-0.1cm}
NGC 4062 & Sc  & 10.02 & 2.78 & 0.04 & pseudo & FD08, W+09 \\ 
\vspace{-0.1cm}
NGC 4236 & Sdm & 8.91 & 2.78  & 0.00 & none & -- \\ 
\vspace{-0.1cm}
NGC 4258 & Sbc & 10.85 & 3.38 & 0.00 & none & --  \\ 
\vspace{-0.1cm}
NGC 4318 & E3 & 9.78 & 2.12 & 0.55 & classical & -- \\
\vspace{-0.1cm}
NGC 4374 & E1 & 11.60 & 3.37 & 0.98 & classical & -- \\
\vspace{-0.1cm}
NGC 4378 & Sa & 11.37 & 3.29 & 0.49  & classical? & -- \\
\vspace{-0.1cm}
NGC 4387 & E4 & 10.17 & 2.19 & 0.65 & classical & -- \\
\vspace{-0.1cm}
NGC 4419 & Sa & 10.32 & 2.59 & 0.11 & pseudo & FD10 \\ 
\vspace{-0.1cm}
NGC 4434 & E1 & 10.32 & 2.24 & 0.77 & classical & --  \\
\vspace{-0.1cm}
NGC 4448 & Sab & 9.96 & 2.51 & 0.22 & pseudo & F+12 \\ 
\vspace{-0.1cm}
NGC 4464 & S0 & 9.85 & 1.74 & 0.52 & classical & -- \\
\vspace{-0.1cm}
NGC 4478 & E2 & 10.39 & 2.25 & 0.79 & classical & -- \\
\vspace{-0.1cm}
NGC 4494 & E1 & 11.00 & 2.85 & 0.80 & classical & -- \\
\vspace{-0.1cm}
NGC 4551 & E3 & 10.14 & 2.13 & 0.72 & classical & -- \\
\vspace{-0.1cm}
NGC 4564 & E5 & 10.44 & 2.78 & 0.48 & classical & FD08, K+11 \\ 
\vspace{-0.1cm}
NGC 4594 & Sa & 11.49 & 3.38 & 0.85 & classical & KK04, FD11 \\ 
\vspace{-0.1cm}
NGC 4605 & Sc & 9.25 & 2.14 & 0.02 & pseudo & FD10 \\ 
\vspace{-0.1cm}
NGC 4697 & E4 & 10.94 & 2.72  & 0.67 & classical & -- \\
\vspace{-0.1cm}
NGC 4698 & Sab & 10.85 & 2.89 & 0.59 & classical? & KB11, F+12 \\ 
\vspace{-0.1cm}
NGC 4736 & Sab & 10.60 & 2.51 & 0.42 & pseudo & KK04, F+12 \\ 
\vspace{-0.1cm}
NGC 4845 & Sab & 10.98 & 3.24 & 0.08 & pseudo? & -- \\
\vspace{-0.1cm}
NGC 5033 & Sc & 10.97 & 3.38  & 0.19 & pseudo & FD10 \\ 
\vspace{-0.1cm}
NGC 5055 & Sbc & 10.92 & 3.19 & 0.07 & pseudo & F+12 \\ 
\vspace{-0.1cm}
NGC 5128 & S0 & 11.21 & 3.01 & 0.72 & classical & -- \\
\vspace{-0.1cm}
NGC 5846 & E1 & 11.40 & 2.84 & 0.96 & classical & -- \\
\vspace{-0.1cm}
NGC 6314 & Sa & 11.31 & 3.07 & 0.74 & classical & N+17 \\ 
\vspace{-0.1cm}
NGC 7171 & Sb & 10.59 & 3.21 & 0.06 & pseudo? & dS+04 \\ 
\vspace{-0.1cm}
NGC 7217 & Sab & 10.96 & 3.00 & 0.27 & classical? & KB11, F+12 \\ 
\vspace{-0.1cm}
NGC 7331 & Sb & 11.17 & 3.16 & 0.32 & classical? & KK04, F+12 \\ 
\vspace{-0.1cm}
NGC 7537 & Sbc & 10.11 & 2.67 & 0.35 & pseudo & B+07 \\ 
\vspace{-0.1cm}
NGC 7541 & Sbc & 10.95 & 3.33 & 0.02 & pseudo? & -- \\
\vspace{-0.1cm}
NGC 7606 & Sb & 11.17 & 3.44 & 0.10 & pseudo? & -- \\
\vspace{-0.1cm}
NGC 7617 & S0 & 10.74 & 2.75 & 0.61 & classical & -- \\
\vspace{-0.1cm}
NGC 7664 & Sc & 10.61 & 2.91 & 0.04 & pseudo & FL+14 \\ 
\vspace{-0.1cm}
IC 467 & Sc & 10.15 & 3.02 & 0.00 & none & -- \\
\vspace{-0.1cm}
UGC 11810 & Sbc & 10.45 & 3.30 & 0.10 & pseudo & FL+14 \\ 
\vspace{-0.1cm}
UGC 12810 & Sc & 11.00 & 3.51 & 0.11 & pseudo?  & -- \\
\vspace{-0.3cm}
\enddata
\tablecomments{
This table is a revision of Tables 3--5 in Paper~1 with the values of $M_\star$,  $j_\star$, and $\beta_\star$ calculated as described in Paper~2.
Galaxies with missing colors or peculiar types are not included here.
Bulge types followed by a question mark are uncertain, as discussed in Section~3.
References for bulge types are abbreviated as follows.
B+07:\ Balcells et al.\ (2007);
C+13:\ Cortesi et al.\ (2013);
dS+04:\ de Souza et al.\ (2004);
FD08:\ Fisher \& Drory (2008);
FD10:\ Fisher \& Drory (2010);
FD11:\ Fisher \& Drory (2011);
F+12:\ Fabricius et al.\ (2012);
FL+14:\ Fern\'andez Lorenzo et al.\ (2014);
GH17:\ Gao \& Ho (2017);
K+11:\ Kormendy et al.\ (2011);
KB11:\ Kormendy \& Bender (2011);
KK04:\ Kormendy \& Kennicutt (2004);
M04:\ M\"ollenhoff (2004);
MA+18:\ M\'endez-Abreu et al.\ (2018);
N+17:\ Neumann et al.\ (2017);
S+18:\ Sweet et al.\ (2018);
W+09:\ Weinzirl et al.\ (2009).}
\end{deluxetable*}

\clearpage
\newpage

\appendix

\section{Errors in \MakeLowercase{\it J}$_{\star}$, $M_{\star}$, {\rm and} $\beta_{\star}$}

The purpose of this appendix is to provide some further insight into both random and systematic errors in the stellar specific angular momentum $j_{\star}$, mass $M_{\star}$, and bulge fraction $\beta_{\star}$. 
We begin by comparing the independent estimates of these quantities by different authors for the 6--10 galaxies in common between our dataset and those of Obreschkow \& Glazebrook (2014) and Posti et al.\ (2018a). 
These are plotted against each other in Figure~4.
Evidently, the correlations between the different estimates are nearly linear, apart from a tendency by Posti et al.\ (2018a) to assign $\beta_\star = 0$ to galaxies with small bulges. 
The mean offsets are $\Delta \log j_{\star} = 0.06$, $\Delta\log M_{\star} = 0.06$, and $\Delta\beta_{\star} = 0.10$, and the one-sample dispersions (i.e., two-sample dispersions divided by $\sqrt2$) about them are $\sigma(\log j_{\star}) = 0.11$, $\sigma(\log M_{\star}) = 0.10$, and $\sigma(\beta_{\star}) = 0.09$.
These results are consistent with our estimates of the total errors $\varepsilon(\log j_{\star})$, $\varepsilon(\log M_{\star})$, and $\varepsilon(\beta_{\star})$ quoted at the end of Section 2.1.

The different methods of disk--bulge decomposition can lead to discrepancies in the derived values of $\beta_\star$, especially for pseudo bulges.
In the method pioneered by Kent (1986), decomposition is based on the fundamental physical distinction between flat (rotation-supported) disks and round (dispersion-supported) bulges.
In the more familiar method, decomposition is based on imposed templates for the surface brightness profiles: exponential for disks and S\'ersic for bulges. 
However, it is important to recognize that there is no fundamental physical justification for imposed templates of exactly these forms, from either cosmology or stellar dynamics.
The different values of $\beta_\star$ returned by the two decomposition methods mostly reflect the fact that real bulges have a variety of 3D shapes and surface brightness profiles, rather than measurement errors. 
Fortunately, both methods usually give similar values of the radial scale of the disk $R_{\rm d}$, typically within $\sim10$\%, and hence similar values of the disk contribution to $j_{\star}$.  (See Section 4.1 of Paper 1 for a more complete discussion of this issue.) 

As noted in Section~2.2, the CALIFA part of the Sweet et al.\ (2018) $j_{\star}$--$M_{\star}$ relation for disk-dominated galaxies has the same exponent ($\alpha \approx 0.6$) as the others plotted in Figure~1 but is higher by a factor of about 2.
We do not know the full reason for this offset, but we have found some clues.
When we estimate $M_\star$ for some of the CALIFA galaxies by our own methods, we usually obtain results within $\sim 0.1$~dex of those adopted by Sweet et al.\ (2018) from Falc\'on-Barroso et al.\ (2017).
For disk-dominated CALIFA galaxies, the estimates of specific angular momentum $j_{\star}$, radial scale $R_{\rm d}$, and rotation velocity $V_{\rm f}$ listed in Table~1 of Sweet et al.\ (2018) are typically related by $j_{\star} \sim 5 R_{\rm d} V_{\rm f}$, i.e., about 2.5 times the value of $j_{\star}$ for an ideal disk with an exponential surface density profile and a flat rotation curve, which is known to be a good approximation for most real disks.
Thus, we strongly suspect that the Sweet et al.\ (2018) estimates of $j_{\star}$ suffer from some systematic error of roughly the amount needed to account for the offset between the CALIFA and the other $j_{\star}$--$M_{\star}$ relations.  

In the process of combining datasets to derive their published $j_{\star}$--$M_{\star}$ relation, Sweet et al.\ (2018) introduced another systematic error.
They made a non-linear rescaling of all our estimates of specific angular momentum of the form $j_{\star} \rightarrow j_{\star}^{1.3}$, based on a claim by Obreschkow \& Glazebrook (2014).
If valid, this would induce a corresponding change $\alpha \rightarrow 1.3\alpha$, hence $\alpha \approx 0.6 \rightarrow 0.8$, in the exponent of the power law $j_{\star} \propto M_{\star}^{\alpha}$. 
This, in turn, would spoil the excellent agreement between the $j_{\star}$--$M_{\star}$ scaling relations for disk-dominated galaxies from our work (Paper~2), Obreschkow \& Glazebrook (2014), and Posti et al.\ (2018a) shown in Figure~1, and therefore can be ruled out on this basis alone. 
Furthermore, the Sweet el al.\ (2014) rescaling of $j_{\star}$ is contradicted by the good agreement between different estimates of $j_{\star}$ for individual disk-dominated galaxies shown in the middle panels of Figure~4.
The only discrepant points here belong to galaxies with significant bulges ($\beta_\star > 1/3$), where the different methods of disk--bulge decomposition and assumptions about bulge rotation matter. The rescaling of $j_{\star}$ is another reason Sweet et al.\ (2018) found a high value of the exponent ($\alpha \approx 1)$ in their combined $j_{\star}$--$M_{\star}$ relation. 

\begin{figure*}
\centering{\includegraphics[width=0.85\textwidth]{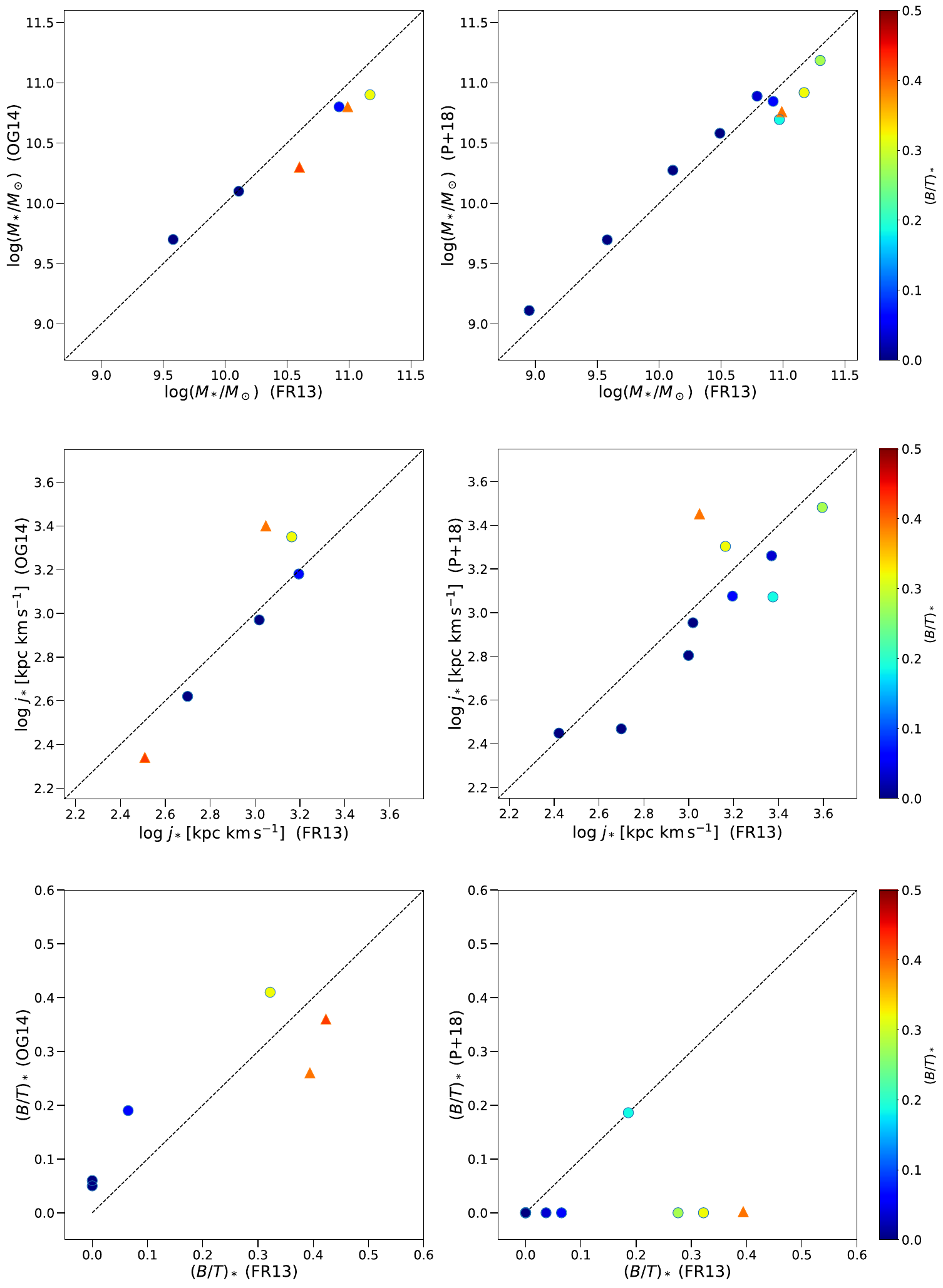}} 
\caption{
Comparison of stellar specific angular momentum $j_{\star}$, stellar mass $M_{\star}$, and
stellar bulge fraction, $\beta_\star \equiv (B/T)_{\star}$, from our work (Paper~2, FR13),
Obreschkow \& Glazebrook (2014, OG14), and Posti et al.\ (2018a, P+18) for the galaxies
in common between these samples.  
The colors and shapes of the plotted symbols indicate the bulge fractions from our dataset,
with circles for $0 \leq \beta_\star < 1/3$, triangles for $1/3 \leq \beta_\star < 2/3$,
and squares for $2/3 \leq \beta_\star \leq 1$. 
Note that the color scale here differs from that in Figure~1.
The dashed diagonal lines indicate the one-to-one relations.
Note that there are no systematic discrepancies between these independent estimates,
apart from a tendency by Posti et al.\ to assign $\beta_\star = 0$ to galaxies with small
bulges. 
}
\end{figure*}

\end{document}